\documentclass[secnumarabic,amssymb,superscriptaddress,floatfix, nobibnotes, aps,prb,onecolumn,longbibliography]{revtex4-1}

\usepackage{amsmath,amssymb,amsfonts}
\usepackage{graphicx}
\usepackage{mathtools}
\usepackage{xcolor}
\usepackage{todonotes}
\usepackage{hyperref}
\usepackage{alphabeta}
\usepackage[normalem]{ulem}

\newcommand{\tblue}[1]{\textcolor{blue}{#1}}

\newcommand{\BFA}{BaFe$_2$As$_2$}
\newcommand{\FeSeS}{FeSe$_{0.9}$S$_{0.1}$}
\newcommand{\SRO}{Sr$_2$RuO$_4$}

\begin{document}
\title{Uniaxial stress effect on the electronic structure of quantum materials}

\author{Na Hyun Jo}
\altaffiliation{nhjo@umich.edu}
\affiliation{Department of Physics, University of Michigan, Ann Arbor, MI 48109, USA}

\author{Elena Gati}
\altaffiliation{elena.gati@cpfs.mpg.de}
\affiliation{Max Planck Institute for Chemical Physics of Solids, 01187 Dresden, Germany}

\author{Heike Pfau}
\altaffiliation{heike.pfau@psu.edu}
\affiliation{Department of Physics, The Pennsylvania State University, University Park, Pennsylvania, 16802 USA}

\date{\today}

\begin{abstract}

Uniaxial stress has proven to be a powerful experimental tuning parameter for effectively controlling lattice, charge, orbital, and spin degrees of freedom in quantum materials. In addition, its ability to manipulate the symmetry of materials has garnered significant attention. Recent technical progress to combine uniaxial stress cells with quantum oscillation and angle-resolved photoemission techniques allowed to study the electronic structure as function of uniaxial stress. This review provides an overview on experimental advancements in methods and examines studies on diverse quantum  materials, encompassing the semimetal WTe$_{2}$, the unconventional superconductor Sr$_{2}$RuO$_{4}$, Fe-based superconductors, and topological materials. 
\end{abstract}

\maketitle


\section{Introduction}

Recent years have seen an tremendous interest in uniaxial stress experiments on quantum materials. While hydrostatic pressure alters electronic orbital overlap in all three spatial dimensions, uniaxial pressure explicitly drives anisotropic changes. Therefore, the material's response can be studied separately for each crystal axis. Often, much larger responses can be obtained with equal amounts of pressures when applied uniaxially. In addition, point group symmetries of crystal structures can be broken.

The perturbation of quantum materials by uniaxial stress has led to a number of important discoveries, for example: It was possible to tune the delicate balance between unconventional superconductivity and competing ordering phenomena in a range of materials. Specifically, the superconducting transition temperature in cuprates was increased by controlling its orthorhombicity \cite{welp_1992,takeshita_2004} and charge order was induced in the underdoped regime by uniaxial stress\,\cite{nakata2022normal,kim_2018}; A dramatic increase in the superconducting $T_c$ with uniaxial stress was also observed in \SRO \cite{Hicks14}, and the material was found to order magnetically at even larger pressures\,\cite{grinenko2021split}; The electronic origin of the rotational symmetry breaking in iron-based superconductors, called nematicity, and many of its properties were discovered by anisotropic strain measurements \cite{chu_2012,boehmer_2022}; Uniaxial pressure also turned out to be a key control parameter for phenomena related to the band-structure topology. For example, transitions between different non-trivial topological phases were induced by uniaxial stress\,\cite{mutch2019evi,zhang2021obs,jo_2023}; Extremely large magneto-elastoresistance was observed from WTe$_{2}$ due to its semi-metallic band structure with different effective masses near the Fermi energy\,\cite{Jo2019}. 

Whereas uniaxial pressure techniques have been available since a long time, they were not intensively used in the study of quantum materials. Recent increased interest was triggered not only by the discovery of various fascinating quantum phenomena but also by new technical developments of stress cells. Apart from standard anvil cells and bending devices, new schemes using thermal contraction\,\cite{Sunko19}, turn-screw mechanisms\,\cite{Tanatar2010}, and in particular piezoelectric devices\,\cite{hicks_2014} were developed. They substantially improve pressure homogeneity and accommodate an ever larger range of experimental probes \cite{Ikeda19,Noad23,Hicks14,Ghosh20}, including thermodynamic, transport, spectroscopy, and scattering techniques in environments with low temperatures, high magnetic fields, or in ultra-high vacuum.

Measurements of the electronic structure under {\it{in-situ}} tunable uniaxial pressure are one of the more recent additions. Key insights into the physics of quantum materials can be obtained from spectral function, electronic band structure, and Fermi surface measurements. They provide information for example about effective masses, Fermi velocities, band gaps due to broken symmetries, and surface electronic states. Quantum oscillation measurements and angle-resolved photoemission spectroscopy (ARPES) are standard probes of the electronic structure of quantum materials \cite{sobota_2021,carrington2011quantum,sebastian2015quantum,hu2019transport}. The technical developments in uniaxial stress cells have therefore been adopted and advanced in recent years in order to combine them with both probes.

Here, we review the current status of uniaxial-stress dependent ARPES and quantum oscillation measurements. We first provide an overview of uniaxial stress devices employed in electronic structure measurements in Chapter \ref{sec:methods}. The following sections present an overview of measurements on several quantum materials. These studies encompass an analysis of strain-induced charge redistribution in WTe$_{2}$ in Chapter \ref{sec:WTe2}, the stress-induced Lifshitz transition in \SRO\ in Chapter \ref{sec:SrRu2O4}, investigation of nematicity in iron-based superconductors in Chapter \ref{sec:FeSC}, and explorations of strain-controlled topological phase transitions in Chapter \ref{sec:topology}.


\section{Uniaxial stress devices for electronic structure measurements}
\label{sec:methods}

In general, the deformation of a solid is described in the elastic regime by $\sigma_{ij} = C_{ijkl} \epsilon_{kl}$, with $\sigma_{ij}$ being the stress tensor, $\epsilon_{kl}$ the strain tensor and $C_{ijkl}$ the elastic stiffness tensors \cite{Luethi05}. All experimental setups described in this review apply a uniaxial stress to a sample, which in turn strains along all crystallographic directions. Thus, uniaxial pressure results in highly anisotropic strains, which can be quantified by the corresponding Poisson's ratio $\nu_{ij}\,=\,-\frac{\epsilon_{jj}}{\epsilon_{ii}}$. Thermal expansion of device components such as sample substrates may induce additional stress components to the sample as function of temperature. Therefore, it is important to quantify strains along all directions for a quantitative understanding of the electronic structure at finite stress and strain.

Despite the crucial significance of investigating changes in electronic structure under the influence of uniaxial stress, these studies have only recently gained momentum after experimental challenges were overcome by sophisticated technical developments. Quantum oscillation experiments require low temperatures and high magnetic fields. Uniaxial stress devices based on piezoelectric stacks have been developed for these challenging conditions in recent years\,\cite{hicks_2014} and are now commercially available.

Implementing \textit{in-situ} tunable uniaxial stress in angle-resolved photoemission spectroscopy (ARPES) experiments presents several additional challenges. Stress devices need to be compatible with the ultra-high vacuum environment; Electric fields distort the trajectories of photoemitted electrons; Most ARPES instruments do not offer electrical contacts at the sample stage; Sample stages are often very small. We present two general types of uniaxial stress devices -- mechanical and piezoelectric devices -- that overcame these challenges and were successfully used in ARPES experiments. In all devices, the sample is mounted across a gap, the size of which is adjusted either mechanically or by piezoelectric stacks. Sample substrates \cite{park_2020} can be used in cases where samples require a large force to cleave or when they tend to bend easily during the cleaving process, or if the sample size is smaller than the size of the gap. As a result, samples of varying dimensions and mechanical properties can be studied in these devices either by measuring them in a free-standing configuration or by supporting them with a substrate. The strain due to applied stress is measured optically with microscope images \cite{Sunko19,cai_2020,Hyun2022}, by x-ray diffraction\cite{zhang2021}, or with strain gauges \cite{pfau_2019_prl,zhang2021,Hyun2022,jo_2023}. 


 \begin{figure*} [ht]
    	\includegraphics[width=4.5 in]{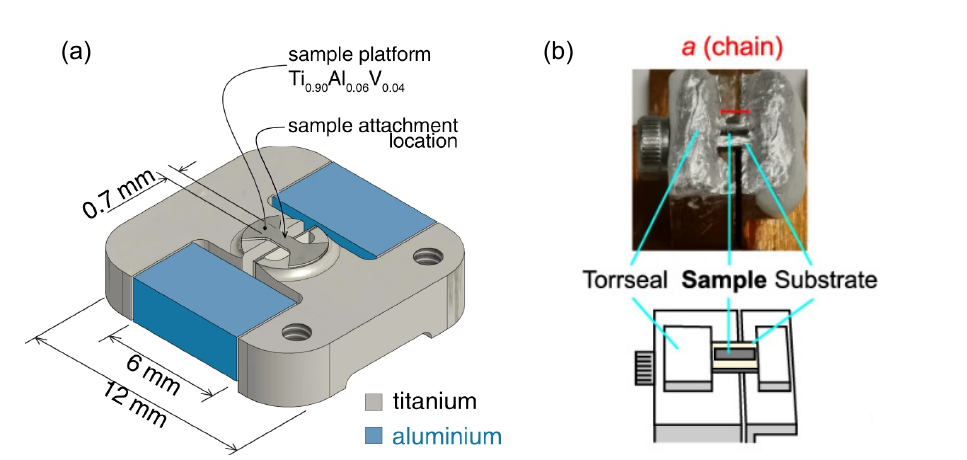}%
    	\caption{Mechanical uniaxial stress devices.
     (a) An illustration of the differential thermal contraction strain device. The thermal contraction of aluminum surpasses that of titanium, resulting in uniaxial compression of the sample platform during the cooling process. (This image is from Sunko $et. al.$ 2019 \cite{Sunko19}. The reference is an Open Access article licensed under a Creative Commons Attribution 4.0 International License.)
     (b) A screw is employed to apply compression or extension to the substrate, consequently affecting the sample attached to it. (This image is from Peng $et. al.$ 2021 \cite{zhang2021obs}. The reference is an Open Access article licensed under a Creative Commons Attribution 4.0 International License.)  
    \label{fig:mechanical}}
\end{figure*}

The first type of mechanical devices leverages the concept of differential thermal contraction. The device by Sunko \textit{et al.}\,\cite{Sunko19} shown in Fig.~\ref{fig:mechanical} (a) employed a combination of titanium (Ti) and aluminum (Al), which leads to uniaxial compression of the sample platform during cooling. Sample strains of up to -\,0.6\,$\%$ at temperatures below 40 K were achieved with this device (see Fig.\,\ref{fig:mechanical}\,(a) for the dimensions of Al and Ti, respectively). 

The second type of mechanical devices uses screw-turn mechanisms that are adjusted {\it{in-situ}} with a wobble stick\,\cite{Kim2011prb, zhang2021, Hyun2022}, which changes the size of the gap (see Fig.~\ref{fig:mechanical} (b)\,\cite{zhang2021}). In other devices, the screw was used to induce bending of a substrate to which the sample is affixed\,\cite{ricco2018,lin2021,nicholson2021}. Since these bending cells induce highly non-uniform strains, they do not qualify as true uniaxial stress cells and thus fall outside the scope of this review article on uniaxial stress.


 \begin{figure*} [ht]
    	\includegraphics[width=5 in]{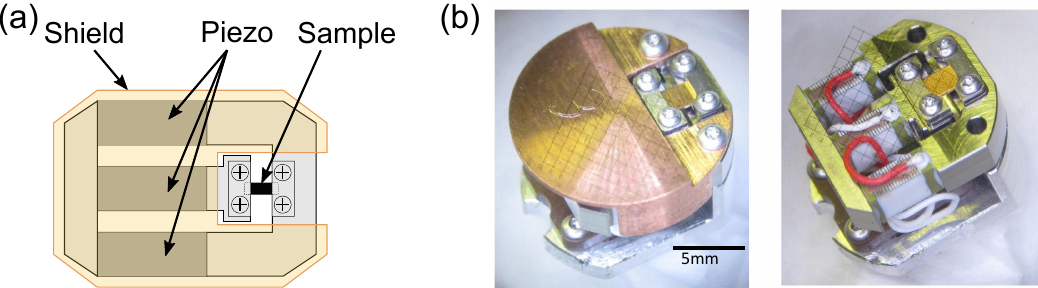}%
    	\caption{Piezoelectric-driven uniaxial stress devices.
     (a) Schematics of the design for use in ARPES. Reproduced with permission.~\cite{pfau_2018} Copyright 2019 by the American Physics Society. (b) Photograph of the device without (right) and with (left) electrical shielding used in Ref.~\cite{pfau_2018,pfau_2019_prl,pfau_2021a_prb} Reproduced with permission.~\cite{pfau_2019_prl,pfau_2021a_prb} Copyright 2019 and Copyright 2021 by the American Physics Society. Reference~\cite{pfau_2018} is an Open Access article licensed under a Creative Commons Attribution 4.0 International License.
    \label{fig:piezo}}
\end{figure*}

Uniaxial stress devices based on piezoelectric stacks have been used in ARPES set-ups that are equipped with electrical contacts on the sample stage \cite{pfau_2018,cai_2020,jo_2023}. These devices offer continuous, {\it{in-situ}}, backlash-free stress tuning. The principle design of all of them is based on Ref.~\cite{hicks_2014} and is adapted to ARPES as shown in Fig.~\ref{fig:piezo}. In particular, an electric shield surrounds the piezoelectric stacks to shield the high voltage from the photoemitted electrons. Different spring-loaded contact designs are employed depending on the specific sample stage.


With these new techniques, diverse quantum materials have been investigated, unveiling exotic physics. In the following sections, we will discuss each case individually.


\section{Charge redistribution in WTe$_{2}$}
\label{sec:WTe2}

As a result of changes in the electronic overlap, the application of uniaxial stress can modify band structures. Stress can influence the curvature of the band, causing changes in the effective mass and the size of the Fermi surface, consequently leading to modifications in carrier density. Changes of the Fermi surface in weakly-correlated systems can be predicted using density-functional theory (DFT) and determined experimentally using quantum oscillation measurements. The study by Jo \textit{et al.} combines these two techniques and showcases the systematic, quantitative tracing of these effects in WTe$_{2}$\,\cite{Jo2019}.

\begin{figure*} [ht]
    	\includegraphics[width=6 in]{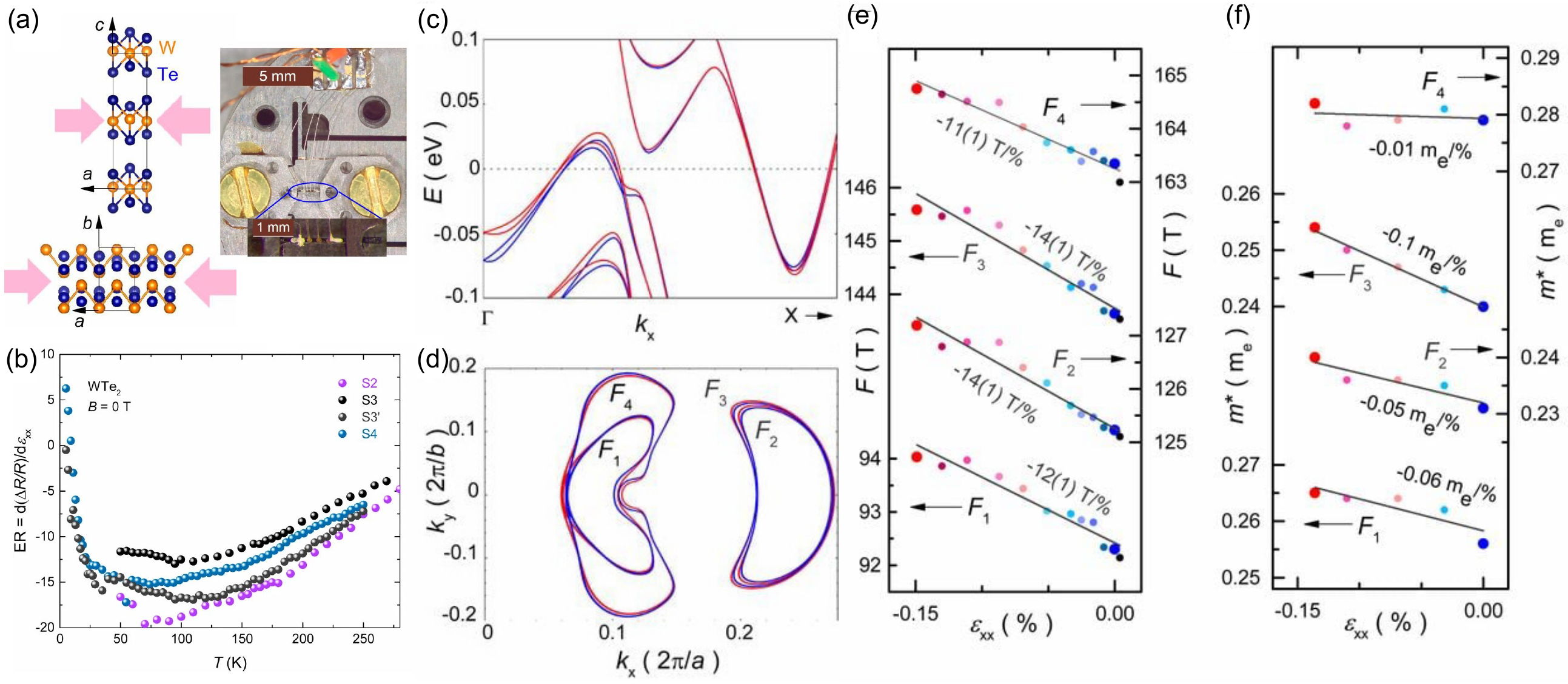}%
    	\caption{Density functional theory (DFT) results and quantum oscillation analysis under strain. (a) Crystal structure of WTe$_{2}$. (Left)  The top picture shows a view on the $ac$ plane to visualize the layered structure along the $c$ axis. The lower pictures shows the $ab$ plane with distorted zig-zag chains of W atoms in the $a$ direction. A single crystal of WTe$_{2}$ is mounted on Razorbill CS100 Cryogenic uniaxial stress cell (Piezoelectric device) (right). The electrical current and mechanical stress were applied along the crystallographic $a$ direction, and the magnetic field was applied along the crystallographic $c$ diretion. (b) Elastoresistance of WTe$_{2}$ was measured in the temperature range of 5\,K to 270\,K, with no applied magnetic field for samples S2, S3, S3' and S4. It predominantly exhibits negative elastoresistance, with a pronounced upturn observed at low temperatures, specifically below 25\,K. (c) Results of DFT band structure calculation along the $\Gamma\,-\,X$ direction without strain ($\epsilon_{xx}$\,=\,0 \,$\%$;  blue) and with strain ($\epsilon_{xx}$\,=\,-0.2\,$\%$; red) applied along the $a$ axis. (d) Strain-induced modification of extremal orbits at $k_{z}$\,=\,0 from DFT calculation. Blue and red lines refer to the same strain as in (c). Fermi surfaces $F_{1}$ and $F_{4}$ correspond to hole bands and $F_{2}$ and $F_{3}$ to electron bands. (e) Shubnikov-de-Haas oscillation frequencies of the 4 extremal orbits $F_{1,2,3,4}$ as a function of strain $\epsilon_{xx}$. The numerical values of the slopes are given in the figure. (f) Effective cyclotron masses of the four extremal orbits $F_{1,2,3,4}$ as a function of strain, with slopes given in the figure. Figure is adapted from Ref. \cite{Jo2019}. Copyright (2019) National Academy of Sciences.
    \label{fig:WTe2}}
\end{figure*}

WTe$_{2}$ is an excellent testbed system to study the influence of uniaxial stress on the electronic structure of quantum materials. It possesses an orthorhombic crystal structure (space group number 31) as shown in Fig.~\ref{fig:WTe2} (a). Consequently, the application of uniaxial stress does not lower the material's crystalline symmetries. While WTe$_{2}$ undergoes a Lifshitz transition as function of temperature at $\approx\,$160\,K\,\cite{Wu2015Lif}, no uniaxial stress-induced phase transitions have been observed so far at low temperatures. Note that a superconducting transition occurred upon the application of large hydrostatic pressure\,\cite{pan2015pressure,kang2015superconductivity}. However, the paper did not indicate a superconducting phase transition down to 2 K with a maximum of -0.15\,$\%$ of uniaxial stress. 

Additionally, the elastoresistance value was found to exceed two at room temperature and exhibited non-monotonic behavior as a function of temperature as shown in Fig.~\ref{fig:WTe2} (b). Elastoresistance describes changes in resistance relative to the applied strain (elastoresistance = $\frac{d(\Delta R/R)}{d \epsilon }$, where R represents resistance and $\epsilon$ represents strain). The elastoresistance of metals is often dominated by changes in geometric factors, resulting in a temperature-independent value of approximately 2. However, few metals exhibit a temperature-dependent, significant elastoresistance value, primarily dominated by the resistivity term, which reflects the intrinsic physical properties of the material. Therefore, the elastoresistance of WTe$_2$ is goverened by strong strain-induced modifications of the electronic structure, rather than by simple changes of the aspect ratio of the sample. The uniaxial stress-dependent study of the electronic structure by Jo \textit{et al.} provided the necessary microscopic insights to understand the magneto- and elastoresistance of WTe$_{2}$.

In this study, the stress was applied along the crystallographic $a$-axis using the stress cell shown in Fig.~\ref{fig:WTe2}(a). Figures\,\ref{fig:WTe2} (c) and (d) depict the results of DFT calculations conducted without (blue) and with (red) a compressive strain of -0.2\,$\%$ along the $a$-axis. They predict two hole bands and two nearly degenerate electron bands to intersect the Fermi energy. The corresponding four Fermi surface pockets are labeled $F_1$ to $F_4$. As function of strain, slight curvature adjustments in the dispersion are discernible in Fig.\,\ref{fig:WTe2} (c), while Fig.\,\ref{fig:WTe2} (d) illustrates an evident increase in pocket sizes.

These findings were confirmed through studies of the Shubnikov-de-Haas quantum oscillations. The Fourier transformation of the experimental data revealed four distinct peaks, each corresponding to the crossing of the four bands at the Fermi energy. Additionally, effective masses were extracted from a fit of the temperature-dependent quantum oscillation amplitude with the Lifshitz-Kosevich theory. The experimental results are shown in Fig.~\ref{fig:WTe2}(e) and (f). The frequencies increase upon application of compressive stress, which indicates an expansion in the extremal cross-sectional area of the Fermi surface. The effective masses exhibit an increase under compressive strain with a slope that is strongly band dependent. All of these observations qualitatively agree well with DFT calculations. However, the exact values do not coincide with those measured in experiments. This discrepancy can be attributed to: 1) DFT typically exhibits relative errors in the lattice constant depending on the choice of exchange-correlation functional, and 2) DFT calculations are conducted at 0 K and do not account for finite-temperature effects. 

The observation of stress-induced changes in band structure provided crucial insights into the unusual transport properties of WTe$_{2}$ that exhibits an exceptionally large elastoresistance with a non-monotonic-temperature-dependence. The increase of the size for all Fermi surfaces with compressive strain implies that electrons redistribute between hole-like and electron-like bands at low temperatures. The larger Fermi surfaces also increase significantly both electron and hole carrier densities. As a result, the low-temperature elastoresistance is exceptionally large and positive as observed experimentally. The temperature dependence of the elastoresistance can be captured qualitatively by a low-energy model, which also takes into account the redistribution of charge carriers to bands within $k_B T$. Especially at high temperatures, this includes a heavy hole band below the Fermi energy that could potentially be observed with ARPES. In general, the transport and electronic structure studies point to a strong connection between the elastic and electronic degrees of freedom in WTe$_{2}$. Due to its small carrier density, the example of WTe$_2$ showcases a novel avenue for manipulating magneto-transport properties through strain in this and related materials.


\section{Stress-driven Lifshitz transitions in Sr$_{2}$RuO$_{4}$} 
\label{sec:SrRu2O4}

The quest to understand the unconventional pairing state of the superconductor SrRu$_2$O$_4$ \cite{Mackenzie03,Mackenzie17,Hicks14,Grinenko21,Pustogow19} has been a driving force behind numerous technical advancements in applying controlled uniaxial pressure to bulk materials in the last decade. This pursuit has culminated in a range of modern pressure devices \cite{Hicks14b,Barber19,Sunko19}. Conversely, the new technical possibilties have faciliated the discovery of entirely new physics in Sr$_2$RuO$_4$ beyond the one of the superconducting state \cite{Sunko19,Li22,Noad23,Yang23}. In the following, we will focus on the electronic structure of normal-state Sr$_2$RuO$_4$ under uniaxial pressure. For further reviews on the various aspects of the physics of Sr$_2$RuO$_4$, particularly at ambient pressure, see Refs.\,\cite{Bergemann03,Mackenzie03,Mackenzie17}.

 \begin{figure*} [ht]
    	\includegraphics[width=6 in]{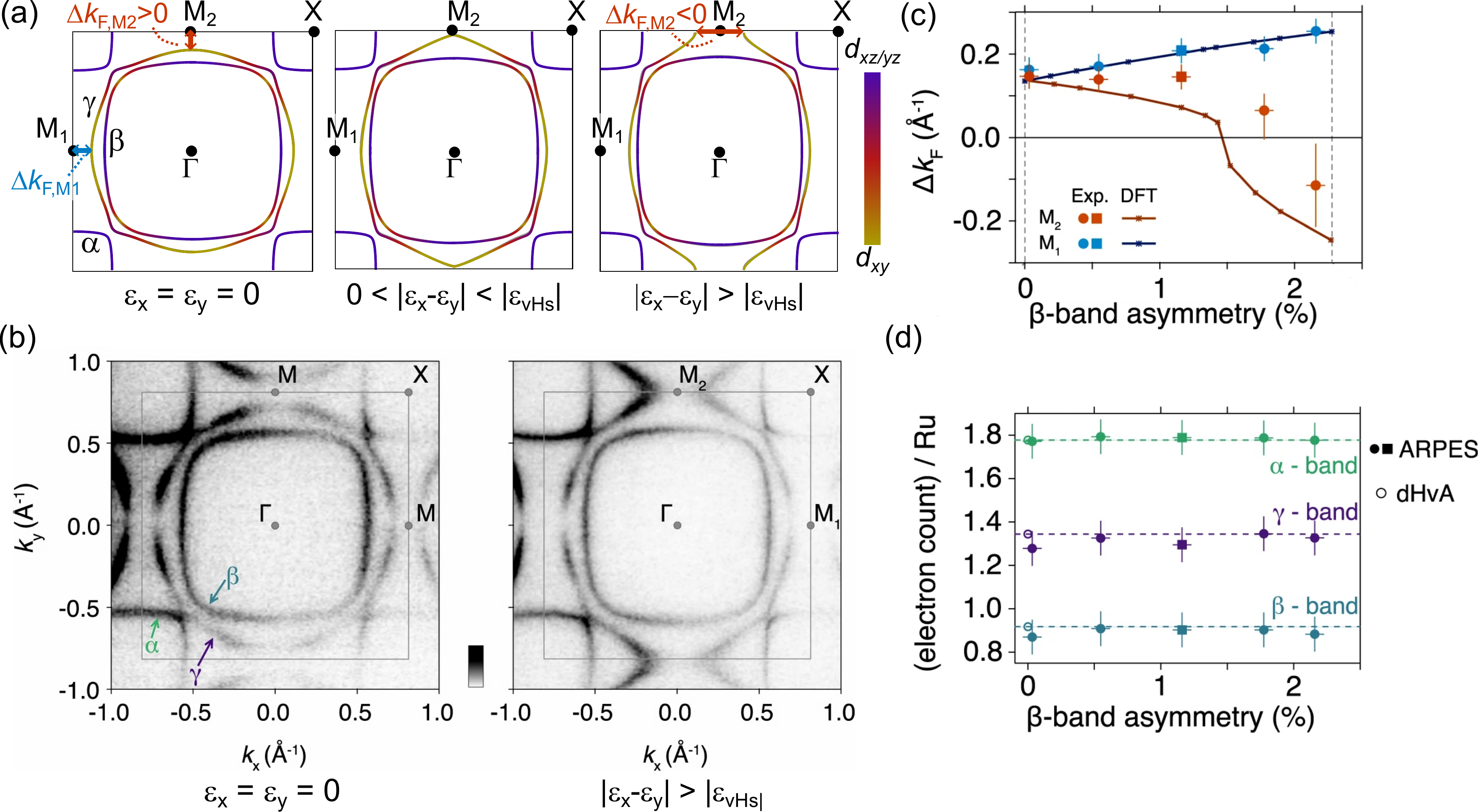}%
    	\caption{Fermi surface of Sr$_2$RuO$_4$ in the unstrained and strained condition (with strain applied along the [1\,0\,0] direction of the tetragonal unit cell). (a) Schematic Fermi surfaces of Sr$_2$RuO$_4$ at zero strain ($\epsilon_x\,=\epsilon_y = \,0$), at intermediate strain (0$<|\epsilon_x-\epsilon_y|<|\epsilon_\textrm{vHs}|$) and at high strains ($|\epsilon_x-\epsilon_y|>|\epsilon_\textrm{vHs}|$). The color shading indicates schematically the orbital character of the eigenstates along the Fermi surface. After Ref.\,\cite{Tamai19,Yang23}. (b) Experimental determination of the Fermi surface through ARPES measurements at zero strain (left) and at high strain (right). (c,d) The anisotropy of the $\gamma$ sheet surface (c) and the electron count as a function of anisotropic strain, which is measured by the $\beta$-band asymmetry. The anisotropy of the $\gamma$ sheet is measured by the parameters $\Delta k_{F,M1}$ and $\Delta k_{F,M2}$, which are defined in the schematic Fermi surfaces in (a). Figures (b), (c), \tblue{and} (d) are adapted from Ref.\,\cite{Sunko19}. The reference~\cite{Sunko19} is an Open Access article licensed under a Creative Commons Attribution 4.0 International License.
    \label{fig:Sr214}}
\end{figure*}

In addition to its unconventional superconductivity, Sr$_2$RuO$_4$ stands out as a correlated metal with a quasi two-dimensional electronic structure which is known in exquisite detail \cite{Tamai19,Mackenzie96,Ohmichi99,Damascelli00,Bergemann00,Bergemann03}. This makes Sr$_2$RuO$_4$ one of the best model systems for comparison with theoretical models related to electronic correlations.

Sr$_2$RuO$_4$ belongs to the class of layered Ruddlesden-Popper series \cite{Ruddlesden58} and exhibits a tetragonal crystal structure. In this structure, the $a$-$b$-layers are formed by corner-sharing RuO$_6$ octahedra. The Fermi surface consists of the $\alpha$, $\beta$ and $\gamma$ bands (see Fig.\,\ref{fig:Sr214}\,(a)), which result from crystal-electric field (CEF) splitting of the 4$d$ manifold of the Ru atom, as well as spin-orbit coupling (SOC). Due to improvements in resolution of ARPES measurements over the years, the effects of correlation-enhanced SOC on the band structure could be visualized and quantified by Tamai \textit{et al.} \cite{Tamai19}. The resulting band structure and the orbital character of the bands can be well fitted by a tight-binding model with five hopping parameters $t$ and a SOC strength $\lambda_\textrm{SOC}$ \cite{Tamai19,Cobo16,Romer19}. This model provides the basis for understanding the strain-induced changes of the electronic structure.

Breaking the in-plane tetragonal symmetry has a remarkably strong effect on the band structure, in particular on the $\gamma$ sheet, as illustrated schematically in Fig.\,\ref{fig:Sr214} (a). When uniaxial pressure is applied along the [1\,0\,0] direction, Sr$_2$RuO$_4$ experiences a compression along the $x$-axis with $\epsilon_x\,<\,0$ and a tension along the $y$-direction, i.e., $\epsilon_y=-\nu_{xy}\epsilon_x>0$ (see Sec.\,\ref{sec:methods}). As a result, upon increasing pressure, the $\gamma$ sheet becomes compressed along the $k_x$ direction, but elongated along the $k_y$ direction. At a sufficiently high strain, the $\gamma$ sheet touches the $M$ point of the Brillouin zone, resulting in a van-Hove singularity (vHs). At this strain value $|\epsilon_x-\epsilon_y|=|\epsilon_\textrm{vHs}|$, the $\gamma$ sheet undergoes a so-called Lifshitz transition \cite{Lifshitz60}, where its Fermi surface topology  changes from a closed configuration (for $|\epsilon_x-\epsilon_y|<|\epsilon_\textrm{vHs}|$) to an open one (for $|\epsilon_x-\epsilon_y|>|\epsilon_\textrm{vHs}|$). 

Naturally, experimental verification of such a Lifshitz transition is not only interesting by itself, but it also gains significance due to the impact on the physical properties of Sr$_2$RuO$_4$ \cite{Li22,Noad23}. Significant experimental results include a more than two-fold increase in the superconducting critical temperature, $T_c$, with increasing compression \cite{Steppke17} to $\epsilon_x,\sim,-0.45$\%. Additionally, deviations from the archetypal Fermi-liquid-type $T^2$ temperature dependence of electrical resistivity were observed in a similar strain range \cite{Barber18}. Experimentally, recent high-precision elastocaloric measurements provided a very detailed thermodynamic map of the phase diagram of Sr$_2$RuO$_4$ as a function of $\epsilon_x$. The thermodynamic identification of the features of the Lifshitz transition \cite{Li22,Noad23} and superconductivity clearly demonstrate that maximum $T_c$ occurs at $\epsilon_\textrm{vHS}$. Presently, it appears likely that this close interrelation is due to the enhanced density of states at the Lifshitz transition and possibly a concomitant increase of electronic correlation strength.

The challenges in probing directly the electronic structure changes at the Lifshitz transition in Sr$_2$RuO$_4$ under large strains were not only related to the specific challenges of ARPES chamber environment, as described in the Sec.\,\ref{sec:methods}, but also related to the elastic properties of Sr$_2$RuO$_4$ \cite{Hicks14b,Noad23,Sunko19}. Owing to its high elastic modulus, it is generally hard to apply large strains. In addition, the elastic limit of Sr$_2$RuO$_4$ at room temperature is quite low, only about $-0.2\,\%$ \cite{Barber19}. This strain is less than the strain needed to reach the vHs. Thus, to probe the Lifshitz transition a device is needed, where some degree of tunability of strain is achieved as the temperature is lowered. This has led Sunko \textit{et al.} to design the mechanical device, based on the concept of differential thermal expansion (see Sec.\,\ref{sec:methods}).

The great success of their approach is clearly visible in Fig.\,\ref{fig:Sr214}\,(b), when comparing the data \cite{Sunko19} taken at $\epsilon_x\,=\,0$ to the data at $|\epsilon_x-\epsilon_y|>|\epsilon_\textrm{vHs}|$. The change of Fermi surface topology of the $\gamma$ sheet from a closed to an open contour becomes clearly visible.

Further quantitative results on the evolution of the electronic structure were inferred from ARPES data \cite{Sunko19} as a function of varying degrees of strain. In such a passive strain device, the variation was achieved through changes of the sample's thickness and corresponding variations in the effective strain transfer from the substrate to the top surface of the sample. Among the key results are the following. First, the $\beta$ sheet responds linearly to strain, whereas the $\gamma$ sheet shows a strongly non-linear change in particular close to $\epsilon_\textrm{vHs}$. This is shown in Fig.\,\ref{fig:Sr214} (c), where the anisotropy of the $\gamma$ sheet, measured by the parameters $\Delta k_{F,M1}$ and $\Delta k_{F,M2}$ (see Fig.\,\ref{fig:Sr214}\,(a)), is plotted against the asymmetry of the $\beta$ sheet, which is taken as a good measure of the anisotropic strain.
Second, the Luttinger count of each of the $\alpha$, $\beta$ and $\gamma$ bands remains unchanged with strain (see Fig.\,\ref{fig:Sr214}\,(d)). This suggests that the Lifshitz transition is solely driven by the distortion of the bands and not by a redistribution of the carriers between the bands.

Tight-binding calculations with hopping parameters $t$, which change linearly with strain, qualitatively confirm the experimental results. The rate d$t$/d$\epsilon$ is taken to be orbital-dependent and results from changes of the Ru-O-Ru bond lengths. Quantitative discrepancies between these calculations and experiments were most pronounced in the evolution of the $\gamma$ sheet along the $k_y$ direction (see Fig.\,\ref{fig:Sr214}\,(c)). It was suggested that electronic correlations are responsible. This notion is supported by the strong change of the Hall coefficient across the Lifshitz transition \cite{Yang23}, which points towards a strong change of electron-electron scattering \cite{Zingl19} and correlation strength in its proximity. Further ARPES measurements under strain will certainly be useful to understand how correlations are precisely altered through the Lifshitz transition.

Finally, the recent study of the surface states of Sr$_2$RuO$_4$ under strain \cite{Damascelli00,Kreisel21} by ARPES provided interesting new results \cite{Abarca23}. In contrast to the bulk, the RuO$_6$ octahedra at the surface of Sr$_2$RuO$_4$ are rotated around the $c$-axis with antiphase on neighboring sites. As a result, the unit cell is enlarged and the Fermi surface significantly reconstructed \cite{veenstra_2013}. The high-quality experimental ARPES data of Abarca Morales \textit{et al.} \cite{Abarca23}, together with their tight-binding modelling, showed that the strain-induced changes of the surface states can be best understood by only considering changes of the Ru-O-Ru bond lengths rather than by changes in the angles. Thus, the surface of Sr$_2$RuO$_4$ by its own forms an interesting reference system to understand the effect of strain on the electronic structure of quantum materials, where the specific arrangement of transition-metal octahedra plays a key role in the emergent physics.


\section{Nematicity in iron-based superconductors}
\label{sec:FeSC}

Nematicity is an electronic instability that breaks rotational symmetry but preserves translational symmetry. It has proven to be a ubiquitous feature of correlated quantum materials and in particular of unconventional superconductors. Nematicity has been studied in great detail in iron-based superconductors (FeSC), which serve as a prototype material class for this instability \cite{paglione_2010,johnston_2010,fisher_2011,fernandes_2014,si_2016,boehmer_2017,boehmer_2022}. In the  FeSCs, nematicty breaks the tetragonal $C_4$ symmetry and yields an order parameter with $B_{2g}$ symmetry. Corresponding anisotropies can be observed in the spin, orbital, and lattice degrees of freedom. 

In many FeSCs, the nematic phase is accompanied by a spin-density wave transition at only slightly lower temperatures. Spin fluctuations were therefore suggested to drive an Ising-spin nematic order as a precursor to stripe magnetism \cite{fang_2008,xu_2008,fernandes_2012}. The discovery of nematicity without a magnetic order in FeSe \cite{mcqueen_2009} promoted the idea of orbital fluctuation as the driving mechanism \cite{lv_2009,lee_2009,chen_2010,onari_2012}. Apart from low-energy descriptions, recent studies also explored the interplay of nematicity and electronic correlations in FeSC \cite{fanfarillo_2017,hu_2018,yu_2018,benfatto_2018,kostin_2018,steffensen_2021,pfau_2021a_prb,pfau_2021b_prb,fanfarillo_2023}. Despite extensive studies of nematicity in the FeSCs, there is so far no consensus on its microscopic origin.

The electronic nematic order changes the crystal structure of FeSCs from tetragonal to orthorhombic due to nemato-elastic coupling. An anisotropic strain of $B_{2g}$ symmetry will therefore induce a finite value of the nematic order parameter at all temperatures. A corresponding linear response measurement returns the nematic susceptibility. The resistivity anisotropy is the most prominent observable that has been used as a proxy for the nematic order parameter. The nematic susceptibility extracted from elastoresistivity measurements follows a Curie-Weiss law in a large temperature range above the nematic transition temperature, $T_\mathrm{nem}$, before it decreases inside the nematic phase \cite{chu_2012}. Apart from resistivity, various other transport, thermodynamic, scattering, and spectroscopic probes have been employed in strain-dependent measurements to extract the nematic susceptibility \cite{boehmer_2014,kissikov_2018,caglieris_2021,sanchez_2021,Occhialini_2023,ikeda_2021,cai_2020}. The temperature-dependence of the susceptibility is qualitatively very similar in almost all of them. However, it is unclear which observables are a true representation of the nematic order parameter since they probe different aspects of the electronic system. Nematic susceptibility measurements across the doping and substitution phase diagrams of FeSCs reveal growing evidence for a nematic quantum critical point.  \cite{palmstrom_2022,Ishida_2022,Worasaran_2021} If it exists, the consequences for the electronic properties are exciting: The diverging nematic fluctuations can lead to non-Fermi liquid behavior where the quasiparticles on the whole Fermi surface become overdamped \cite{metzner_2003}. At the same time the fluctuations can lead to strong, long-range interactions that promote high-temperature superconductivity \cite{yamase_2013,maier_2014,metlitski_2015,lederer_2017,labat_2017}.

The continuous development and sophistication of strain tuning capabilities will be an important factor to resolve the current questions in the field of nematicity in the FeSCs. The recent implementation of uniaxial stress capabilities in ARPES experiments allows to measure the effect of a $B_{2g}$ strain on the electronic structure. A corresponding linear response measurement provides a momentum, orbital, and energy-resolved nematic susceptibility. We will describe in detail how the electronic structure changes inside the nematic state in Section \ref{sec:FeSC_ElectronicStrucure}. This will guide the discussion of ARPES studies under uniaxial pressure in Sections \ref{sec:FeSC_Ba122} and \ref{sec:FeSC_FeSe}.

\subsection{Electronic structure across the nematic phase transition}
\label{sec:FeSC_ElectronicStrucure}

\begin{figure*} [ht]
    	\includegraphics[width=6 in]{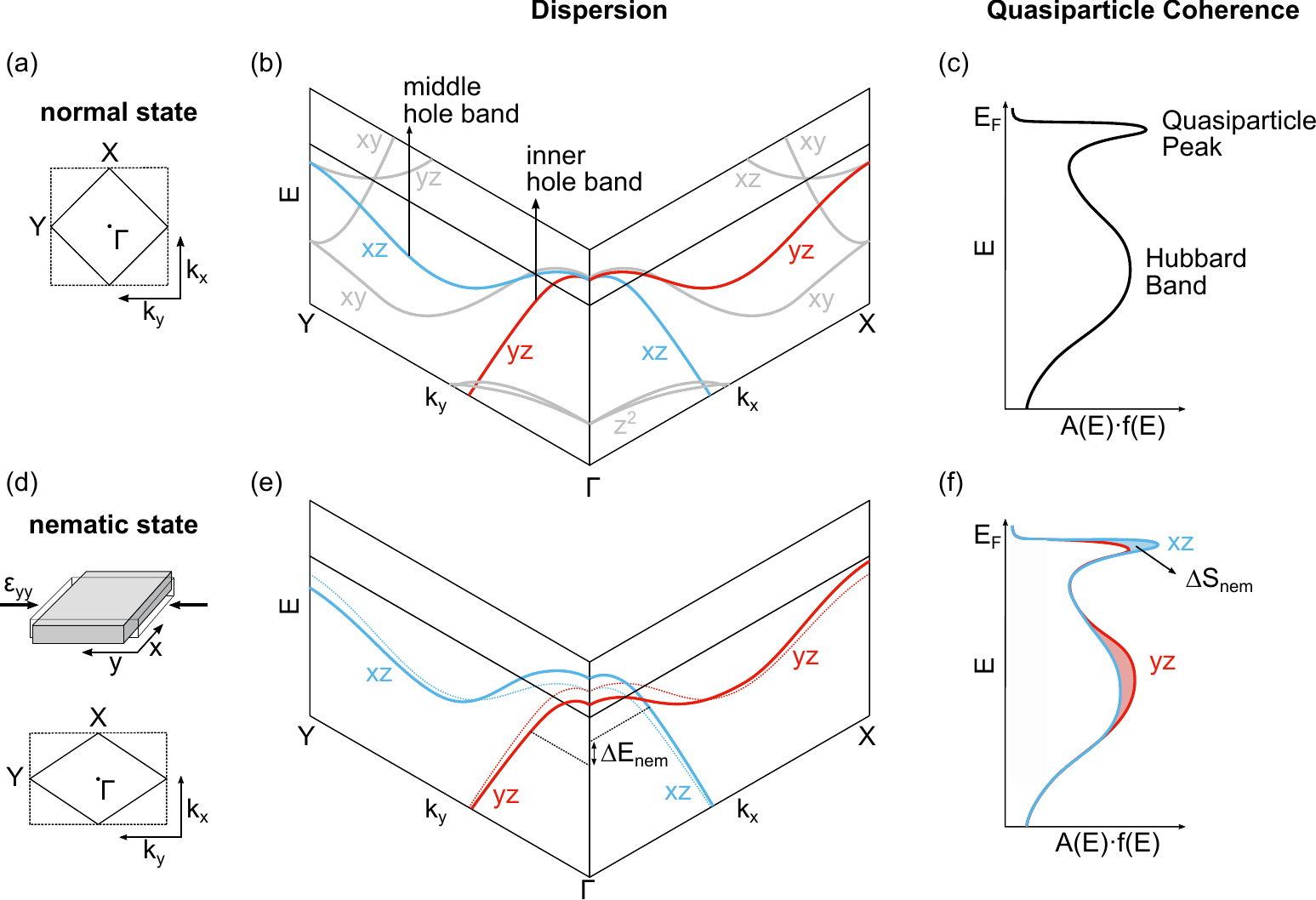}%
    	\caption{Effects of nematicity on the spectral function. (a) The 1Fe (dashed) and 2Fe (solid) Brillouin zone in the tetragonal state. (b) Sketch of the low energy band structure. While specifics such as band width and chemical potential vary between different FeSCs, the main features are consistent across the family. (c) Over a larger energy range, the spectral function $A$ shows a splitting of the $3d$ weight into a quasiparticle peak and Hubbard bands. $A$ is multiplied by a Fermi-Dirac distribution $f$. (d) Brillouin zone in the nematic state. The sketch of the sample indicates the corresponding deformation due to nematicity or due to uniaxial compression along $y$. (e) Change in dispersion of the two $d_{xz}$ and $d_{yz}$ hole bands in the nematic state. Dashed lines indicate dispersion in normal state for comparison. $\Delta E_\mathrm{nem}$ is the nematic band splitting. (f) Sketch of the observed anisotropy of quasiparticle coherence in nematic state, that can be characterized by $\Delta S_\mathrm{nem}$ \cite{pfau_2021a_prb,pfau_2021b_prb}.  (a),(b),(d) and (e) have been adapted with permission \cite{pfau_2021a_prb}. Copyright 2021 by the American Physics Society.
    \label{fig:nematicity}}
\end{figure*}

The Fe $3d$ orbitals form the low-energy electronic structure and Fermi surface of FeSCs as shown in Fig.~\ref{fig:nematicity}(b). The orbitals are split by the tetragonal crystal electric field, which leads to a degeneracy between the $d_{xz}$ and $d_{yz}$ orbitals. This degeneracy is lifted when the material enters the nematic state, which leads to two key signatures in the spectral function that have been observed experimentally with ARPES. 

1) The dispersion of the bands that predominantly contain $d_{xz}$ and $d_{yz}$ orbital character changes inside the nematic state. It can be characterized by a nematic band splitting $\Delta E_\mathrm{nem}$, which is determined from the binding energy difference between the $d_{xz}$ and $d_{yz}$ bands as shown in Fig.~\ref{fig:nematicity}(e). This splitting has been characterized in detail with ARPES for both hole and electron bands \cite{yi_2011_pnas,jensen_2011,kim_2011,yi_2012_njp,nakayama_2014,watson_2015,suzuki_2015,fedorov_2016,zhang_2016,watson_2017_njp,pfau_2019_prb,fedorov_2019_prb,pfau_2019_prl,yi_2019_prx}. We highlight the behavior of the hole bands in Fig.~\ref{fig:nematicity} because they are the focus of the uniaxial stress experiments published so far.

2) Electronic correlations due to the interplay of Coulomb interaction and Hund's rule coupling split the spectral function into a coherent quasiparticle peak and incoherent Hubbard bands as show in Fig.~\ref{fig:nematicity}(c). Hubbard bands have experimentally been observed in FeSe \cite{evtushinsky_2016_arxiv,watson_2017_prb}. The ratio of spectral weight in the Hubbard bands and the quasiparticle peak characterizes quasiparticle coherence or equivalently the degree of electronic localization. Recent ARPES experiments showed that the $d_{xz}$ and $d_{yz}$ orbital have a different degree of quasiparticle coherence inside the nematic phase as sketched in Fig~\ref{fig:nematicity}(f) \cite{pfau_2021a_prb,pfau_2021b_prb}. This effect can be characterized by the spectral weight difference $\Delta S_\mathrm{nem}$. In general, correlation effects such as enhanced effective masses and the suppression of quasiparticle coherence are strongly orbital dependent in FeSCs \cite{yi_2017_npj}. The observation of the spectral weight anisotropy inside the nematic state reflects this orbital differentiation since the degeneracy of the $d_{xz}$ and $d_{yz}$ is lifted inside the nematic state.

\subsection{Universality of nematic response with and without magnetic order from strain-dependent ARPES}
\label{sec:FeSC_Ba122}

The absence of magnetic order made FeSe the ideal candidate to study the effects of nematicity on the electronic structure because band folding due to spin-density wave (SDW) order did not obstruct its signatures. The detailed momentum dependence of $\Delta E_\mathrm{nem}$ and the complex dispersion response at the electron bands  was mapped out in detail in FeSe \cite{nakayama_2014,shimojima_2014,watson_2015,suzuki_2015,fanfarillo_2016,fedorov_2016,watson_2016,zhang_2016,watson_2017_njp,yi_2019_prx,pfau_2019_prl}. The anisotropy of quasiparticle coherence inside the nematic phase was revealed in FeSe as well \cite{pfau_2021b_prb}. Comparable measurements of the pristine nematic state for magnetic FeSCs are restricted to a very small temperature window between the nematic and magnetic transition. They are, however, highly desirable to probe whether the microscopic mechanism of nematicity in magnetic and non-magnetic FeSCs is different.

\begin{figure*} [ht]
    	\includegraphics[width=\textwidth]{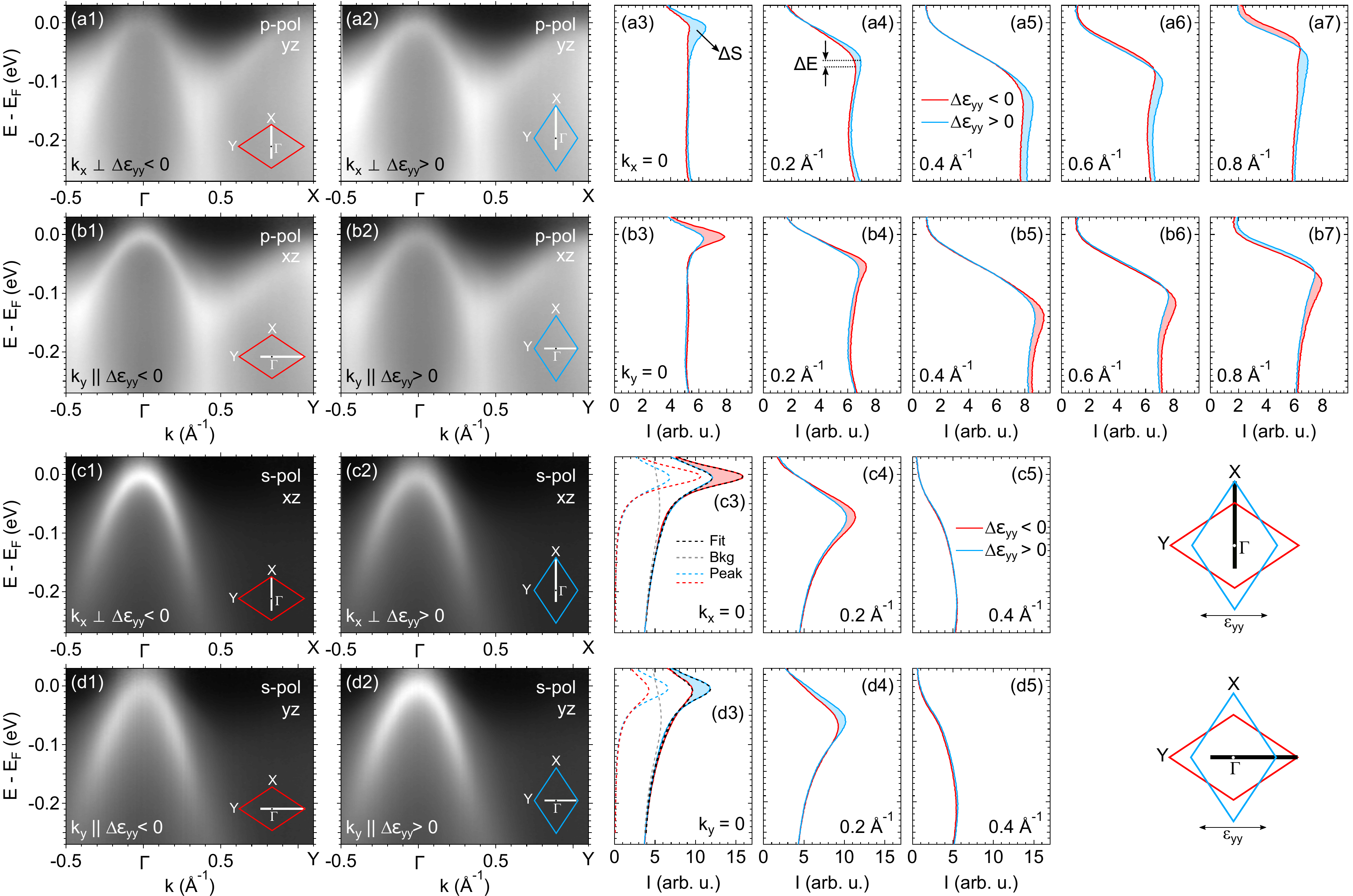}%
    	\caption{ARPES spectra on \BFA~under uniaxial pressure at 160 K. (a) Spectra taken with p-polarized light, which highlights the middle hole band. It has predominantly $d_{yz}$ character along $k_x$. (a1) is taken with compressive stress applied along $y$ while the sample is tensioned for the spectrum in (a2). (a3)-(a7) show energy distribution curves (EDCs) extracted from both spectra at the indicated momentum. Each panel compares the compressed with the tensioned state. Stress induced changes of the band dispersion $\Delta E$ are obtained from the maximum of the EDCs. The change in quasiparticle spectral weight is indicated by $\Delta S$. (b) same as in (a) but along $k_y$, which leads to photoemisison predominantly from the $d_{xz}$ orbital. (c,d) same as (a,b) but with s-polarized light, which photoemits electrons from the inner hole band. Sketches in the bottom right indicate the measurement geometries with the Brillioun zone for compressive (red) and tensile (blue) stress $\epsilon_{yy}$ along $y$. Figures are reproduced with permission Ref.~\cite{pfau_2021a_prb}. Copyright 2021 by the American Physics Society.
    \label{fig:Ba122_spectra}}
\end{figure*}

To overcome the interference from SDW order, recent ARPES studies on \BFA~used in-situ tunable uniaxial stress (see Section \ref{sec:methods}) along the Fe-Fe bond direction applied at a temperature above $T_\mathrm{nem}=140$\,K using a piezoelectric device  \cite{pfau_2019_prl,pfau_2021a_prb}. By inducing a finite nematic order parameter at all temperatures through $B_{2g}$ strains, this approach allows to study the response of the electronic structure to nematicity without interference of the SDW order.

\begin{figure*} [ht]
    	\includegraphics[width=3 in]{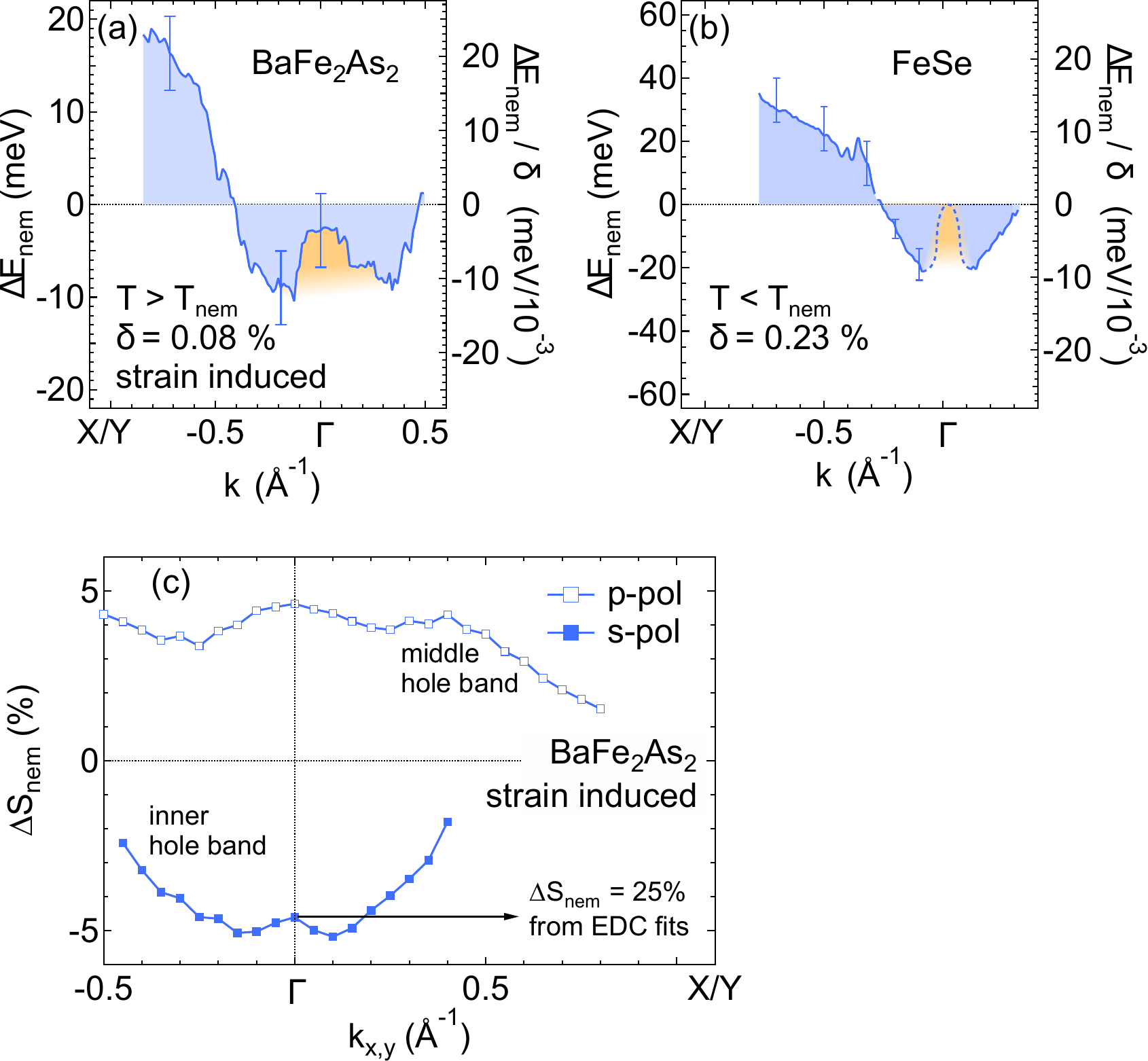}%
    	\caption{(a) Uniaxial-stress induced nematic band splitting of the middle hole band in \BFA. $\delta$ is the orthorhombicity extracted from the measured strain. (b) Equivalent data on FeSe but inside the nematic state. Both materials show the same momentum dependence and overall scale of the nematic band splitting. (c) Uniaxial-stress induced anisotropic quasiparticle coherence in \BFA. It is normalized to the total spectral weight in the measured energy window. The magnitude of the spectral weight response extracted from fits shown in Fig.~\ref{fig:Ba122_spectra}(c3) is indicated at $\mathrm{\Gamma}$. (a),(b) Reproduced with permission~\cite{pfau_2019_prl}. Copyright 2019, American Physical Society. (c) Reproduced with permission~\cite{pfau_2021a_prb}. Copyright 2021, American Physical Society
    \label{fig:Ba122_results}}
\end{figure*}

Figure \ref{fig:Ba122_spectra} shows a comprehensive overview of the ARPES results. The uniaxial stress direction is labeled $y$ without loss of generality. Photoemission matrix elements are employed to selectively probe either $d_{xz}$ or $d_{yz}$ orbital character of the inner and the middle hole band. The two spectra for each configuration are taken at a compressed and tensioned state of the sample, respectively. The ARPES data show clear uniaxial stress-induced changes of the electronic structure, that can be identified in the energy distribution curves (EDCs). First, a shift in binding energy $\Delta E$ was obtained from the peak maxima (Fig.~\ref{fig:Ba122_spectra}(a4)). Second, a change of the quasiparticle spectral weight $\Delta S$ was obtained from the difference in ARPES intensity (Fig.~\ref{fig:Ba122_spectra}(a3)). Both quantities were evaluated along $k_x$ and along $k_y$. The associated nematic band splitting and anisotropic quasiparticle coherence is calculated from the antisymmetric component of these observables (Eqn.1) and shown in Fig.~\ref{fig:Ba122_results}.
\begin{equation}
\begin{aligned}
    \Delta E_\mathrm{nem} &= \frac{\Delta E(k_y) - \Delta E(k_x)}{2}\\
    \Delta S_\mathrm{nem} &= \frac{\Delta S(k_y) - \Delta S(k_x)}{2}
\end{aligned}
\end{equation}
The raw data in Fig.~\ref{fig:Ba122_spectra} already show that the response to uniaxial pressure is almost entirely antisymmetric, i.e. it follows a $B_{2g}$ symmetry and therefore corresponds to changes in the spectral function due to a finite nematic order parameter. 

Figure \ref{fig:Ba122_results}(a) shows the momentum-dependent nematic band splitting in \BFA, which can directly be compared to the one obtain for FeSe inside the nematic state (Fig.~\ref{fig:Ba122_results}(b)). Both show a sign change between $\mathrm{\Gamma}$ and the Brillouin zone corner. The complex response around $\mathrm{\Gamma}$ is due to an interplay of spin-orbit coupling and nematic band splitting, which has been described in detail in Ref.~\cite{pfau_2019_prl}. The magnitude of the splitting, normalized by the corresponding orthorhombicity $\delta$, is very similar as well. Figure \ref{fig:Ba122_results}(c) shows the momentum dependence of the strain-induced nematic spectral weight response for the inner and the middle hole band. $\Delta S_\mathrm{nem}$ has the same sign in \BFA~as in FeSe \cite{pfau_2021b_prb}, i.e. the $d_{xz}$ orbital becomes more coherent than the $d_{yz}$. The size of the response, once normalized to the orthorhombicity, is also on the same order of magnitude. These results suggest that the microscopic mechanism behind nematicity in the non-magnetic FeSe and the magnetic \BFA~is identical.

\subsection{Linear strain response of the band dispersion -- Nematic susceptibility}
\label{sec:FeSC_FeSe}

The nematic band splitting $\Delta E_\mathrm{nem}$, which is non-zero only below nematic phase transition temperature $T_\mathrm{nem}$, can be interpreted as an order parameter. Its linear strain response and the corresponding nematic susceptibility will detect nematic fluctuations in the charge channel and in an orbital, momentum, and energy-resolved fashion.
This knowledge can greatly contribute to current discussions about the microscopic origin of nematicity and about the influence of nematic fluctuations on the strange metal regime. The interpretation of the band shift as an order parameter neglects spin-orbit interaction, which is on the order of a few tens of meV in FeSCs. It leads in principle to a non-linear relationship between the nematic order parameter and the measured band shifts \cite{fernandes_2014,pfau_2019_prl}. 

\begin{figure*} [ht]
    	\includegraphics[width=\textwidth]{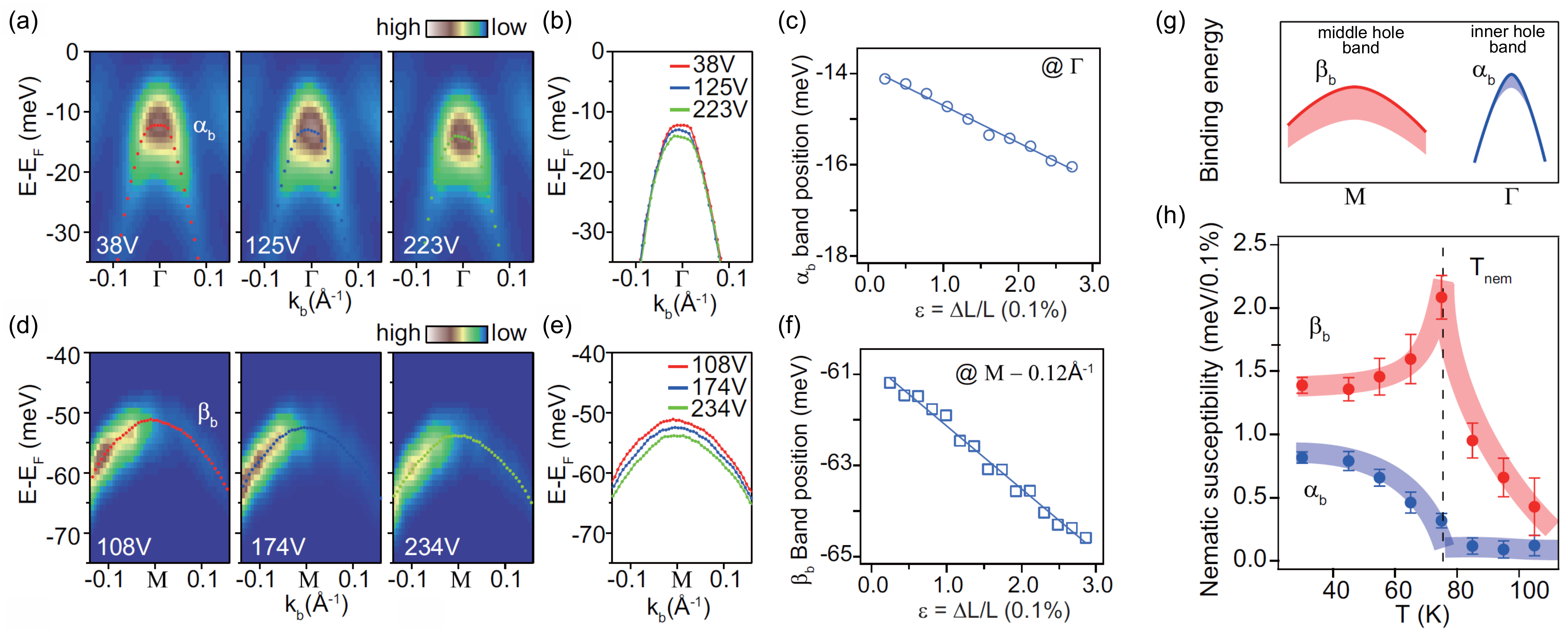}%
    	\caption{Linear strain response of the dispersion and corresponding nematic susceptibility in \FeSeS at 30 K below the nematic transition temperature. (a) ARPES spectra of the inner hole band around the Brillouin center taken at different voltages of the piezo-driven stress device. The points indicate the band position obtained from peak fitting. They are replotted in (b) to illustrate the voltage dependence of the dispersion. (c) The binding energy at $\mathrm{\Gamma}$ follows a linear behavior as function of the strain component $\epsilon$ along the pressure direction. (d)-(f) same as in (a)-(c) but for the middle hole band measured around the Brillouin zone corner M. (g) Sketch of the measured dispersion and the band shift due to uniaxial stress. (h) Temperature dependence of the nematic susceptibility extracted from the linear response of the band shift. Figures are reproduced with permission~\cite{cai_2020}. Copyright 2020, American Physical Society.
    \label{fig:FeSe}}
\end{figure*}

Cai \textit{et al.} performed the first experiments to determine the temperature-dependent nematic susceptibility with ARPES. They measured the strain-response of the dispersion in \FeSeS~($T_\mathrm{nem} = 65$ K) using a piezoelectric device \cite{cai_2020}. Selected ARPES spectra of the inner hole band ($\alpha$) and the middle hole band ($\beta$) as function of the voltage applied to the piezoelectric stacks are shown in Fig.~\ref{fig:FeSe}(a) and (d). The inner hole band is studied around the Brillouin zone center $\mathrm{\Gamma}$ while the middle hole band was studied around the Brillouin zone corner M. Dispersions are extracted from intensity maxima and plotted as function of voltage in Fig.~\ref{fig:FeSe}(b) and (e), respectively. A clear change of the dispersion is detected that varies continuously with uniaxial stress. 

The data was quantitatively evaluated as function of strain $\epsilon$, which was measured parallel to the stress direction using microscope images of the gap between the two sample mounting surfaces. The data in Fig.~\ref{fig:FeSe} (c,f) show that the binding energy is linear as function of $\epsilon$ below $T_\mathrm{nem}$ (Fig.~\ref{fig:FeSe})(c,f). This is in contrast to the hysteresis detected in the spectral weight response due to domain redistribution (not shown). Cai \textit{et al.} also find a linear response for higher temperatures and across $T_\mathrm{nem}$. Non-linear contributions due to spin-orbit coupling were not observed.

Using the observed linear response, Cai \textit{et al.} calculated the nematic susceptibility as function of temperature as shown in Fig.~\ref{fig:FeSe}(h). Interestingly, the susceptibility is different for the two different bands and momenta. For the middle hole band ($\beta$) at a momentum close to M, the nematic susceptibility has a peak at $T_\mathrm{nem}$ and resembles a Curie-Weiss behavior. The temperature dependence is qualitatively similar to that from other probes such as resistivity or Raman spectroscopy \cite{chu_2012,gallais_2016}. In contrast, the peak at $T_\mathrm{nem}$ is absent in the susceptibility derived from the inner hole band ($\alpha$) at $\mathrm{\Gamma}$. The origin of this behavior is unclear so far and possible scenarios include the presence of multiple order parameters at the Brillouin zone center and boundary \cite{fernandes_2014}, or the influence of spin-orbit coupling, which affects the dispersion particularly at $\mathrm{\Gamma}$ \cite{fernandes_2014,pfau_2019_prl}. Additionally, the role of the $d_{xy}$ orbital in the formation of nematic order is still debated \cite{watson_2016,yi_2019_prx,rhodes_2022}.


\section{Strain-induced topological phase transitions}
\label{sec:topology}

Topologically non-trivial materials have attracted significant attention due to their robust, dissipationless electronic states. Controlling these exotic states of matter is crucial for their application. Therefore, topological phase transitions from trivial to non-trivial topological states as well as between different non-trivial states have emerged as a significant research field.

One of the most notable methods to manipulate topological states involves breaking time-reversal symmetry. This can be achieved through the application of a magnetic field or doping with magnetic elements. An early attempt involves the random distribution of magnetic elements on the surface of a topological material\,\cite{chen2010massive}. However, this approach lacks reversibility. A different method employs magnetic topological materials\,\cite{bernevig2022progress}; however, this approach limits the usable temperature and magnetic field range.

An alternative strategy involves the application of uniaxial stress. Uniaxial stress not only offers reversibility but also provides a means to easily engineer and fine-tune topological phase transitions. This makes it a valuable tool in manipulating the exotic electronic states of topological materials. 
Topological phase transitions driven by uniaxial and multi-axial strains have been predicted in various materials including Bi$_{2}$Se$_{3}$, Cd$_{3}$As$_{2}$, ZrTe$_{5}$, HfTe$_{5}$, TaSe$_{3}$, LnPn (Ln = Ce, Pr, Sm, Gd, Yb; Pn=Sb, Bi) and more\,\cite{Young2011_prb,Wang2013,Weng2014prl,Nie_2018_prb,duan2018tunable}. 
Among them, three different bulk topological materials have been studied with ARPES under uniaxial stress: ZrTe$_{5}$, HfTe$_{5}$, and TaSe$_{3}$. In the following, we will present the results from ARPES studies, which directly visualize the degree of tunability and control of the topological properties with uniaxial stress.

\begin{figure*} [ht]
    	\includegraphics[width=5.5 in]{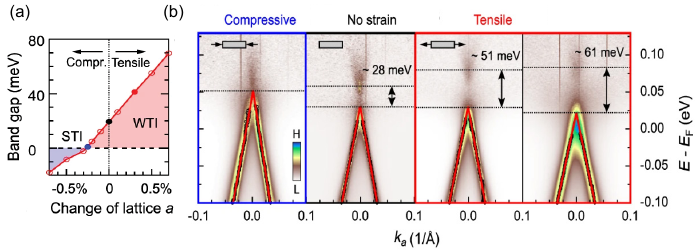}%
    	\caption{Topological phase transitions in transition-metal pentatellurides. (a) Calculated phase diagram of ZrTe$_{5}$ with different lattice constants (strain)\,\cite{zhang2021obs}. Blue, black, and red solid circles roughly indicate the experimental values in (b). (b) The ARPES results on ZrTe$_{5}$ are shown. The bulk band gap changes in size with compressive and tensile strain. With compressive strain, the gap is (nearly) closed, reaching a Dirac semimetal state. With tensile strain, the band gap becomes larger, stabilizing the WTI state. The data are taken with $p$-polarized photons and normalized by their density of states (DOS). The black markers are extracted from the momentum distribution curve (MDC) peaks, and the red solid lines are the fitting results of the black markers.\cite{zhang2021obs} The reference~\cite{zhang2021obs} is an Open Access article licensed under a Creative Commons Attribution 4.0 International License.}
    \label{fig:TPT}
\end{figure*}

The transition metal pentatellurides ZrTe$_{5}$ and HfTe$_{5}$ have an orthorhombic crystal structure (Space group number 63, $Cmcm$). They are layered materials due to van-der-Waals bonding between Te along the crystallographic $b$ direction. In the late 70's and early 80's, these materials received a lot of attention due to peculiar behavior in the resistivity\,\cite{okada1980giant,skelton1982giant,DiSalvo1981}. Their topological behavior and in particular the prospect of strain tuning topological phase transitions brought them back onto the map of condensed matter research\,\cite{li2016chiral,Chen2015,Chen20152,Zheng2016,Shen2017,Wu2016prx,Li2016prl,Manzoni2015,zhang2017nc,Luca2016,Manzoni2016,liu2018, Nair_et_al:2018}. The band inversion that is responsible for the bulk band gap and the non-trivial topological state is not due to spin-orbit coupling. Instead it is driven by the special nonsymmorphic space group. Specifically, the band inversion at $\Gamma$ occurs between the zigzag chain Te$^{z}$-$P_{x}$ and the prism chain dimer Te$^{z}$-$P_{y}$ states. Both Te$^{z}$-$P_{x}$ and Te$^{z}$-$P_{y}$ are initially fourfold degenerate due to the presence of four equivalent Te atoms. However, strong interchain covalent bonding causes each state to split into two, and weak interchain coupling further divides them into singly degenerate states. Consequently, the band inversion occurs between the bonding Te$^{d}$ states with m$_{xz}$=1 and the antibonding Te$^{z}$ states with $p$=-1, resulting in one odd parity state. This change in occupation alters the total parity of the occupied states at $\Gamma$, leading to the emergence of the topological state. Hydrostatic pressure does not alter the topological state as long as the interlayer coupling is not strong enough to reverse the band ordering\,\cite{Weng2014prl}. In contrast, according to DFT results, uniaxial stress along the crystallographic $a$ axis modifies the energy gap effectively and allows to tune between an STI and WTI state as this material is in the boundary\,\cite{mutch2019evi}. The corresponding phase diagram is shown in Fig.~\ref{fig:TPT}(a).

Zhang \textit{et al.} performed ARPES measurements under uniaxial stress to visualize the predicted topological phase transition on ZrTe$_{5}$\,\cite{zhang2021obs}. They utilized a mechanical stress device as depicted in Fig.\,\ref{fig:mechanical} (b). The band structure measured under different stress is shown in Fig.~\ref{fig:TPT} (b). The unstrained sample exhibits a bandgap of approximately 28 meV. Upon the application of tensile stress, the bandgap expands and hence the WTI state is stabilized. Under compressive stress, the gap size diminishes and completely disappears, which marks the theoretically-predicted topological phase transition into a Dirac semimetal (DSM) state. Calculations suggest that further compression will lead to a STI phase\,\cite{Weng2014prl, zhang2021obs}. The isostructural HfTe$_{5}$ was predicted to undergo the same topological phase transition as ZrTe$_{5}$. Jo \textit{et al.} performed APRES measurements under in-situ tunable stress on HfTe$_{5}$ using a piezoelectric device\,\cite{jo2023effects}. The results also indicate a topological phase transition from a WTI to a STI with applied compressive stress. Furthermore, this research highlights that the self-energy of electronic states are strongly impacted by the presence of topological surface states.

\begin{figure*} [hbt]
    	\includegraphics[width=5 in]{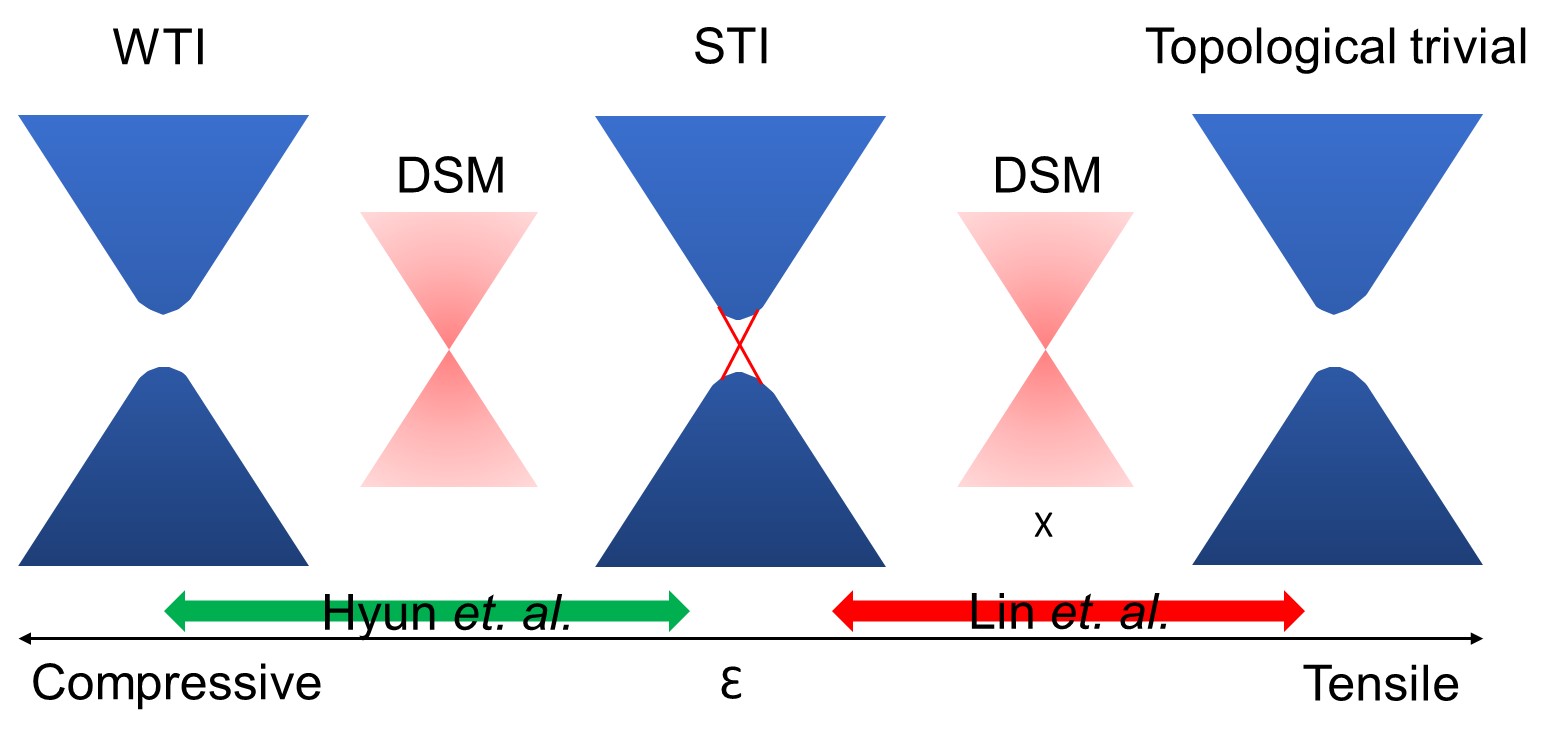}%
    	\caption{A schematic topological phase diagram of TaSe$_{2}$ is presented. The green arrow marks the regime studied in reference,\cite{Hyun2022}, while the red arrow marks the regime explored in reference,\cite{lin2021}. However, in reference \cite{lin2021}, the Dirac semimetal (DSM) state between STI and trivial insulator was not observed.}
    \label{fig:TaSe}
\end{figure*}

TaSe$_{3}$ is a superconductor with a transition temperature of 2\,K \cite{sambongi_1977}. At the same time, it was predicted to host non-trivial topological states \cite{Nie_2018_prb}. The possibility of topological superconductivity and Majorana physics renders TaSe$_{3}$ a particularly exciting material for strain tuning of topological properties. TaSe$_{3}$ has a monoclinic, quasi one-dimensional crystal structure with chains along the $b$-axis that are coupled by van-der-Waals interactions. The topological nature is due to a band inversion of Ta $d$ states and Se $p$ states. Spin-orbit interaction leads to a gap opening at the band crossing points.

Initial calculations predicted that TaSe$_{3}$ undergoes topological transitions as function of strain perpendicular to the chains along the $a$ and $c$ axis \cite{Nie_2018_prb}. Subsequent ARPES studies combined with DFT calculations by Lin \textit{et al.} and Hyun \textit{et al.} studied topological transitions for strains parallel to the chains along $b$ \cite{lin2021, Hyun2022}. Both studies predicted a phase transition sequence from a WTI through a STI towards topological trivial states but differ in their assignment of the zero-stress topological state (see Fig.\,\ref{fig:TaSe})). This difference was attributed to internal strains from Se vacancies \cite{Hyun2022}. The pioneering study from Lin \textit{et al.} used a bending device to apply stress to the sample. The subsequent study by Hyun \textit{et al.} employed a mechanical uniaxial stress device that is actuated by a screw as described in Section \ref{sec:methods}. Both ARPES measurements confirm stress-induced changes of the topological properties in TaSe$_{3}$ as predicted by their DFT calculations.

The initial experiments have been very successful in showcasing the efficacy of strain-tuning in topological phase transitions. These achievements lay the groundwork for a comprehensive understanding of topological phase transitions in the future.


\section{Conclusions}

A key property of quantum materials is their tunability by non-thermal parameters. Changing material properties by in-situ tuning parameters such as electric and magnetic fields or pressure are of particular importance. They circumvent disorder effects induced by chemical doping and substitution and they ease technical application. In the past decade, uniaxial stress tuning uncovered a host of new phenomena and has emerged as a versatile tool in the study of quantum materials. The increased interest was to a large extent driven by technological developments. This is also reflected in the electronic structure measurements under uniaxial pressure that we present here. In particular, the requirements for the photoelectron emission and detection severely limited the implementation of non-thermal tuning in ARPES so far. The adaptation of uniaxial stress cells for photoemission allows for the first time to measure ARPES as function of in-situ tunable stress. There is a push to expand the range of tuning parameters for ARPES even further for example by adding   magnetic field tuning \cite{huang_2023_rsi,ryu_2023_jes}.
We have discussed how these ARPES measurements, together with quantum oscillation studies, under uniaxial pressure contribute important new insights into fields, such as unconventional superconductivity, correlated electron physics, and topological properties. The examples discussed here demonstrate the importance of these techniques in addressing outstanding questions in the quantum-materials field. Furthermore, these studies not only address existing questions but also pave the way for novel research avenues in quantum materials, in which electronic structure studies under tunable stress will play a crucial role.

Currently, the primary limitations arise from the amount of stress that can be applied to the system. Specifically, when utilizing substrates, they have a constrained elastic response. Looking forward, if we could apply greater uniaxial stress, that would significantly enhance our research capabilities. Another avenue worth pursuing is to improve the spatial resolution via nano-ARPES. Investigating the tuning of strain in two-dimensional materials and their heterostructures could yield fascinating results. Additionally, we could explore new possibilities by broadening the spectrum of stress tensors, including shear forces such as twisting.

\section{acknowledgements}
 
NHJ acknowledges support from National Science Foundation (NSF) through NSF CAREER grant under Award No. DMR-2337535. HP is supported by the U.S. Department of Energy, Office of Science, Office of Basic Energy Sciences, Materials Sciences and Engineering Division, under Award Number DE-SC0024135. EG thanks the Max-Planck-Society for support.

\pagebreak
\section{References}
%

\end{document}